\documentclass{PoS}

\title{Baryon Number Correlations in Heavy Ion Collisions}

\ShortTitle{Baryon Number Correlations in Heavy Ion Collisions}

\author{\speaker{Apoorva Patel}\\
        Centre for High Energy Physics and\\
        Supercomputer Education and Research Centre,\\
        Indian Institute of Science, Bangalore 560012, India\\
        E-mail: \email{adpatel@cts.iisc.ernet.in}}

\abstract{
The cross-over region of the quark-gluon plasma (QGP) created in heavy
ion collisions is influenced by the nearby deconfinement, chiral and
baryon condensation phase transitions. A characteristic signature of
the deconfinement transition in this region can be inferred using the
flux tube model, which is dual to the Polyakov loop description and
which offers a visual picture of what happens during the transition.
The three-point (anti)vertices of a flux tube network lead to formation
of (anti)baryons upon hadronisation. Since there is no fundamental
interaction associated with the baryon number, correlations in the
baryon number distribution at the last scattering surface directly
reflect the preceding pattern of the flux tube vertices in the QGP.
An alternating pattern of vertices and antivertices would lead to an
oscillatory signal in the two-point baryon number correlations, under
the experimental conditions prevalent in heavy ion collisions at RHIC
and LHC. The strength of the oscillations is a measure of the
flexibility of the QGP.}

\FullConference{The 30th International Symposium on Lattice Field Theory\\
		 June 24 - 29,  2012\\
		 Cairns, Australia}

\begin{document}
\input{axodraw.sty}

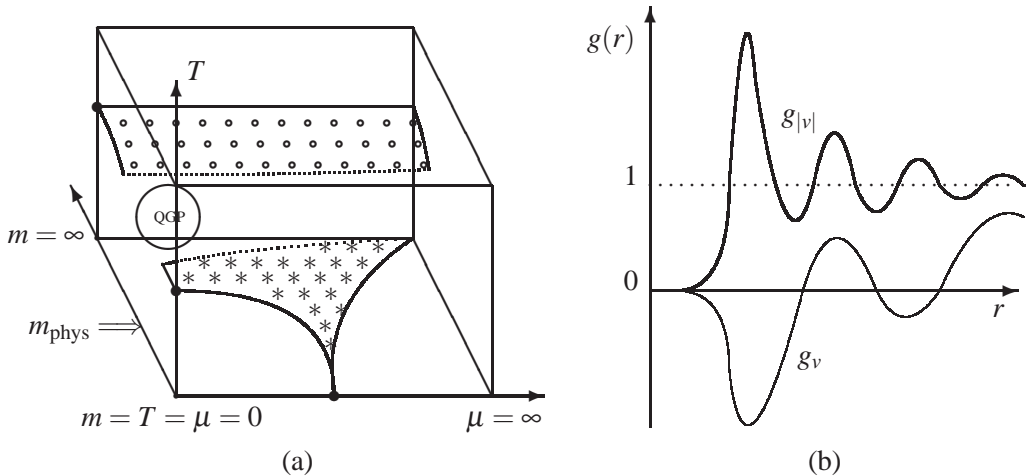
\begin{figure}[b]
\begin{center}{
\setlength{\unitlength}{0.7mm}
\begin{picture}(200,80)
  \thicklines
  \put(30,10){\vector(1,0){70}}
  \put(30,10){\vector(0,1){60}}
  \put(30,10){\vector(-1,2){20}}
  \put(30,50){\line(1,0){60}}
  \put(15,40){\line(1,0){60}}
  \put(15,80){\line(1,0){60}}
  \put(15,40){\line(0,1){40}}
  \put(75,40){\line(0,1){40}}
  \put(90,10){\line(0,1){40}}
  \put(30,50){\line(-1,2){15}}
  \put(90,50){\line(-1,2){15}}
  \put(90,10){\line(-1,2){15}}

  \put(15,65){\circle*{2}}
  \put(30,30){\circle*{2}}
  \put(60,10){\circle*{2}}
  \put(30,10){\circle*{1}}
  \put(15,40){\circle*{1}}

  \put(15,65){\line(1,0){60}}
  \qbezier{(15,65),(18,60),(20,52)}
  \qbezier{(75,65),(77,60),(78,53)}
  \qbezier[60]{(20,52),(50,52),(78,53)}
  \multiput(20,62)(5,0){12}{\circle{1}}
  \multiput(21,58)(5,0){12}{\circle{1}}
  \multiput(22,54)(5,0){12}{\circle{1}}

  \put(30,30){\line(-1,2){2.6}}
  \qbezier[40]{(27.5,35),(40,38),(75,40)}
  \qbezier{(30,30),(60,30),(60,10)}
  \qbezier{(60,10),(57,28),(75,40)}
  \multiput(57,37)(5,0){3}{\makebox(0,0)[bl]{$\ast$}}
  \multiput(34,34)(5,0){7}{\makebox(0,0)[bl]{$\ast$}}
  \multiput(31,31)(5,0){7}{\makebox(0,0)[bl]{$\ast$}}
  \multiput(48,28)(5,0){3}{\makebox(0,0)[bl]{$\ast$}}
  \multiput(55,25)(5,0){2}{\makebox(0,0)[bl]{$\ast$}}
  \put(57,22){\makebox(0,0)[bl]{$\ast$}}
  \put(58,18.5){\makebox(0,0)[bl]{$\ast$}}

  \put(85,3){\makebox(0,0)[bl]{$\mu=\infty$}}
  \put(32,70){\makebox(0,0)[bl]{$T$}}
  \put(-2,39){\makebox(0,0)[bl]{$m=\infty$}}
  \put(12,3){\makebox(0,0)[bl]{$m=T=\mu=0$}}
  \put(2,20){\makebox(0,0)[bl]{$m_{\rm phys}\Longrightarrow$}}
  \put(28.5,44){\circle{12}} \put(25.7,42.5){\makebox(0,0)[bl]{\tiny QGP}}

  \thicklines
  \put(120,30){\vector(1,0){70}}
  \put(120,4){\vector(0,1){80}}

  \qbezier{(125,30),(135,30),(135,50)}
  \qbezier{(135,50),(138,95),(140,70)}
  \qbezier{(140,70),(146,30),(151,50)}
  \qbezier{(151,50),(155,70),(159,50)}
  \qbezier{(159,50),(163,40),(167,50)}
  \qbezier{(167,50),(171,60),(175,50)}
  \qbezier{(175,50),(179,45),(183,50)}
  \qbezier{(183,50),(187,54),(191,50)}

  \thinlines
  \qbezier{(125,30),(135,30),(135,15)}
  \qbezier{(135,15),(138,-12),(149,30)}
  \qbezier{(149,30),(155,50),(163,30)}
  \qbezier{(163,30),(168,20),(175,30)}
  \qbezier{(175,30),(183,48),(191,44)}

  \put(185,25){\makebox(0,0)[bl]{$r$}}
  \put(115,30){\makebox(0,0)[bl]{$0$}}
  \put(115,50){\makebox(0,0)[bl]{$1$}}
  \multiput(120,50)(2,0){35}{\circle*{0.5}}
  \put(108,75){\makebox(0,0)[bl]{$g(r)$}}
  \put(145,60){\makebox(0,0)[bl]{$g_{|v|}$}}
  \put(148,15){\makebox(0,0)[bl]{$g_{v}$}}

  \put(50,-5){\makebox(0,0)[bl]{(a)}}
  \put(150,-5){\makebox(0,0)[bl]{(b)}}
\end{picture}
}\end{center}
\caption{(a) Schematic description of the QCD phase structure in the $m-T-\mu$
space. The first-order transition surfaces are shown shaded, and the critical
lines are shown dotted. The value of $m$ corresponding to the real world QCD,
and the QGP region accessible in heavy-ion collisions, are pointed out.
\hfil\break
(b) Schematic representation of the anticipated baryon number two-point
correlation functions, $g_{|v|}(r)$ (thick line) and $g_v(r)$ (thin line).
The former is similar to that for objects with hard-core repulsion.
The latter is for a percolating flux tube network where vertices and
antivertices alternate.}
\label{fig_one}
\end{figure}

The study of collective non-perturbative phenomena in QCD at high
temperatures and/or at large chemical potentials is a highly active area
of research: experimentally through heavy ion collisions \cite{expt},
theoretically through phenomenological models \cite{pheno},
and computationally through lattice QCD simulations \cite{latt}.
Consider QCD with $N=3$ colours and $N_f$ degenerate quark flavours of
mass $m$, at temperature $T$ and quark chemical potential $\mu$.
The phase structure of this theory is depicted in Fig.\ref{fig_one}(a)
in a schematic manner. Three corners of the phase structure are
well-understood because of the exact symmetries present there:
(a) the finite temperature deconfinement phase transition at $m=\infty$,
governed by the global $Z_3$ centre symmetry of the Polyakov loop,
(b) the finite temperature chiral phase transition at $m=0=\mu$, governed
by the flavour $SU(N_f)_L \otimes SU(N_f)_R$ symmetry, and
(c) the baryon condensation phase transition at $m=0=T$, when $\mu$ crosses
the constituent quark mass.
These end-points are first-order phase transitions, and phase transition
surfaces extend inwards from them. Phenomenological and numerical studies
show that the surfaces end in critical lines, and there is no phase transition
for the physical values of the quark masses as $T$ is varied (unless $\mu$ is
sufficiently large). Nevertheless, the three nearby transitions cause various
QCD properties to change rapidly in the cross-over region. To enhance our
understanding of QCD, it is important to construct observables that can be
extracted from the experimental data and that can highlight the dynamical
features of deconfinement, chiral symmetry restoration and baryon condensation.

In this article, I describe how the deconfinement phase transition leaves
a signature in two-point baryon number correlations in the angular distribution
of the hadrons produced in heavy ion collisions. The collisions at RHIC and
LHC create a high energy state of QCD matter, which upon cooling produces
$O(1000)$ hadrons. These hadrons are dominated by pions, but also include
$O(100)$ baryons and antibaryons. The particle numbers are large enough
for one to analyse their distributions in position, momenta and energy
(instead of only looking at individual particle properties). For the
statistical distributions to be meaningful, they have to arise from some
reasonably equilibrated state. Experimental data suggest that such an
approximately thermalised quark-gluon plasma (QGP) is created, with
energy density $\epsilon \simeq 1{\rm GeV/fm}^3$ and temperature
$T_{\rm cr} \simeq 175{\rm MeV}$, in the fireball of central collisions
and produces moderate $p_T$ hadrons. An important limitation of the
experiments is that they detect only charged hadrons moving in
sufficiently transverse directions.

The experimentally observed multiplicities and distributions of various
particles have been fitted to hadron resonance gas models, resulting in
estimates of the ``chemical freeze-out" temperature (where inelastic
scattering stops) and the ``kinetic freeze-out" temperature (where elastic
scattering stops) \cite{heinz}. The stage is now set to go beyond the
average properties and investigate the two-point correlations in the
distributions. A useful analogy is the study of the cosmic microwave
background radiation (CMBR), where first the black body spectrum was
determined, and then temperature fluctuations were detected at the level
of $\Delta T \simeq 10^{-5} T$ with characteristic angular correlations
\cite{weinberg}. The ``last scattering surface" in the evolution of the
CMBR is analogous to the ``kinetic freeze-out" stage of the QCD fireball.
The patterns that can be observed there would be an invaluable help in
understanding the QGP dynamics that preceded it.

In what follows, I first present a flux tube model providing a physical
picture of the deconfinement process \cite{patel1}, and then point out
that under suitable conditions the scenario predicts a specific two-point
baryon number correlation signal \cite{patel2}, illustrated in
Fig.\ref{fig_one}(b). The challenge for theorists is to quantify this
signal as much as possible, and the challenge for experimentalists is
to detect it.

\section{Flux Tube Model Phenomenology}

The flux tube model of QCD is motivated by the dual superconductor
description of linear colour confinement \cite{ripka}, where condensation
of colour magnetic charges restricts colour electric fields to vortex-like
configurations. The model is phenomenologically quite successful, and is
manifest in lattice QCD results as the area law for large Wilson loops.
A characteristic property of the flux tube is its energy per unit length,
i.e. the string tension $\sigma$. Other than that, the flux tube has a
finite width $w$ and a persistence length $a$, both of order
$\Lambda_{QCD}^{-1}$.

\begin{figure}[b]
\vspace{-6mm}
\epsfysize=5truecm
\centerline{\epsfbox{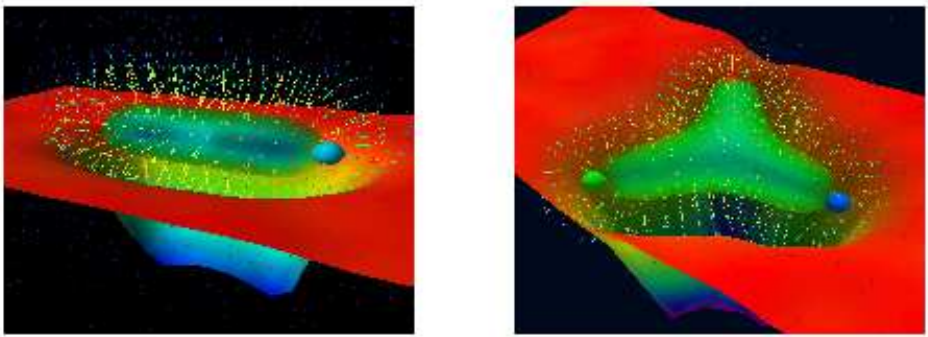}}
\vspace{-4mm}
\caption{Gluon flux tube configurations between static (anti)quark sources
determined by lattice QCD simulations \cite{leinweber}. A meson is on the
left and a baryon is on the right.}
\vspace{-2mm}
\label{fig_two}
\end{figure}

The flux tubes have to obey the constraint of Gauss's law. So they terminate
only on quarks, and interact only at $3$-point vertices. These two features
represent the invariant tensors $\delta_{ab}$ and $\epsilon_{abc}$ used to
describe the meson and the baryon wavefunctions, and are illustrated by the
lattice QCD simulation results displayed in Fig.\ref{fig_two}. Other
multi-quark hadron states are phenomenologically not prominent, and so all
other interactions among the flux tubes are ignored in the model.

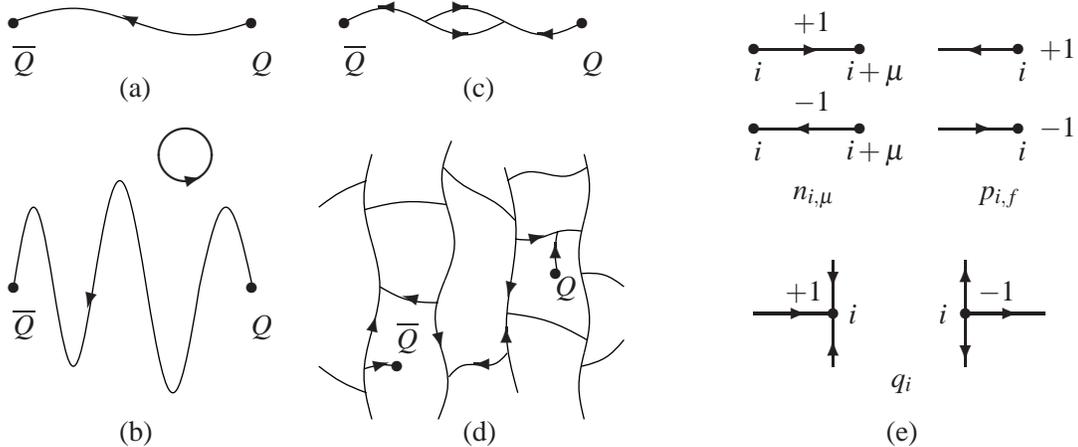
\begin{figure}[b]
\begin{center}{
\setlength{\unitlength}{1pt}
\begin{picture}(400,151)
  \thicklines
  \put(50,120){\makebox(0,0)[bl]{(a)}}
  \Photon(10,150)(100,150){5}{1}
  \put(55,151){\vector(-2,1){4}}
  \put(10,150){\circle*{4}} \put(100,150){\circle*{4}}
  \put(10,130){\makebox(0,0)[bl]{$\overline{Q}$}}
  \put(100,130){\makebox(0,0)[bl]{$Q$}}

  \put(180,120){\makebox(0,0)[bl]{(c)}}
  \Photon(135,150)(195,150){5}{1}
  \Photon(165,150)(225,150){5}{1}
  \put(150,156){\vector(-1,0){2}} \put(210,146){\vector(-1,0){2}}
  \put(180,156){\vector( 1,0){2}} \put(180,146){\vector( 1,0){2}}
  \put(135,150){\circle*{4}} \put(225,150){\circle*{4}}
  \put(135,130){\makebox(0,0)[bl]{$\overline{Q}$}}
  \put(225,130){\makebox(0,0)[bl]{$Q$}}

  \put(50,-10){\makebox(0,0)[bl]{(b)}}
  \put(75,100){\circle{20}}
  \put(77,90.5){\vector(4,1){4}}
  \Photon(10,50)(40,50){30}{1}
  \Photon(40,50)(80,50){40}{1}
  \Photon(80,50)(100,50){30}{0.5}
  \put(40,50){\vector(-1,-4){2}}
  \put(10,50){\circle*{4}} \put(100,50){\circle*{4}}
  \put(10,30){\makebox(0,0)[bl]{$\overline{Q}$}}
  \put(100,30){\makebox(0,0)[bl]{$Q$}}

  \put(180,-10){\makebox(0,0)[bl]{(d)}}
  \Photon(145,0)(145,100){3}{1.5} \Photon(170,0)(175,105){-3}{2}
  \Photon(200,10)(195,95){3}{1}   \Photon(225,0)(225,100){2}{2}
  \Photon(125,80)(143,90){1}{0.5} \Photon(125,20)(143,10){-1}{0.5}
  \Photon(142,79)(173,81){2}{0.5} \Photon(148,50)(170,44){-2}{0.5}
  \Photon(142,19)(155,20){1}{0.5}
  \Photon(172,95)(199,75){-1}{1}  \Photon(174,16)(196,25){2}{1}
  \Photon(196,90)(226,95){2}{1}   \Photon(199,68)(224,71){-1}{1}
  \Photon(196,40)(226,30){2}{0.5} \Photon(215,71)(215,55){-1}{0.5}
  \Photon(224,55)(240,50){2}{0.5} \Photon(224,20)(240,30){-2}{0.5}
  \put(150,21){\vector(4,1){2}}   \put(145,31){\vector(1,4){2}}
  \put(159,46){\vector(-4,1){2}}  \put(170,35){\vector(1,-4){2}}
  \put(186,21){\vector(-1,0){2}}  \put(196.5,33){\vector(0,1){2}}
  \put(199,54){\vector(-1,-4){2}} \put(209,69){\vector(4,1){2}}
  \put(215,65){\vector(0,1){2}}
  \put(155,20){\circle*{4}} \put(215,55){\circle*{4}}
  \put(155,25){\makebox(0,0)[bl]{$\overline{Q}$}}
  \put(215,45){\makebox(0,0)[bl]{$Q$}}

  \thicklines
  \put(290,110){\line(1,0){40}}          \put(290,140){\line(1,0){40}}
  \put(290,110){\circle*{4}}             \put(290,140){\circle*{4}}
  \put(330,110){\circle*{4}}             \put(330,140){\circle*{4}}
  \put(315,110){\vector(-1,0){10}}       \put(305,140){\vector(1,0){10}}
  \put(290,98){\makebox(0,0)[bl]{$i$}}   \put(290,128){\makebox(0,0)[bl]{$i$}}
  \put(325,96){\makebox(0,0)[bl]{$i+\mu$}}
  \put(325,126){\makebox(0,0)[bl]{$i+\mu$}}
  \put(305,116){\makebox(0,0)[bl]{$-1$}} \put(305,146){\makebox(0,0)[bl]{$+1$}}

  \put(360,110){\line(1,0){30}}          \put(360,140){\line(1,0){30}}
  \put(390,110){\circle*{4}}             \put(390,140){\circle*{4}}
  \put(380,140){\vector(-1,0){10}}       \put(370,110){\vector(1,0){10}}
  \put(398,106){\makebox(0,0)[bl]{$-1$}} \put(398,136){\makebox(0,0)[bl]{$+1$}}
  \put(390,98){\makebox(0,0)[bl]{$i$}}   \put(390,128){\makebox(0,0)[bl]{$i$}}

  \put(290,40){\line(1,0){30}}       \put(370,40){\line(1,0){30}}
  \put(320,40){\circle*{4}}          \put(370,40){\circle*{4}}
  \put(300,40){\vector(1,0){10}}     \put(380,40){\vector(1,0){10}}
  \put(302,44){\makebox(0,0)[bl]{$+1$}} \put(375,44){\makebox(0,0)[bl]{$-1$}}
  \put(326,36){\makebox(0,0)[bl]{$i$}}  \put(360,36){\makebox(0,0)[bl]{$i$}}
  \put(320,20){\line(0,1){40}}       \put(370,20){\line(0,1){40}}
  \put(320,20){\vector(0,1){10}}     \put(370,32){\vector(0,-1){10}}
  \put(320,60){\vector(0,-1){10}}    \put(370,48){\vector(0,1){10}}

  \put(305,80){\makebox(0,0)[bl]{$n_{i,\mu}$}}
  \put(375,80){\makebox(0,0)[bl]{$p_{i,f}$}}
  \put(342,10){\makebox(0,0)[bl]{$q_i$}}
  \put(340,-10){\makebox(0,0)[bl]{(e)}}
\end{picture}
}\end{center}
\caption{(a)-(d) depict possible flux tube configurations connecting a static
quark-antiquark pair. Increasing temperature takes (a) to (b) and (c) to (d),
while creation of baryonic vertices takes (a) to (c) and (b) to (d).
(e) lists the link and the site variables used for formulating the flux tube
model on a lattice.}
\label{fig_three}
\end{figure}

The finite temperature behavior of the model is governed by competition
between the energy and the entropy of the flux tube configurations. At low
temperatures, energy wins, keeping down the total length of the flux tubes.
At high temperatures, entropy dominates, producing elaborate structures of
the flux tubes all over the space. As the temperature is increased the flux
tubes oscillate more, and also produce more vertices. Some of the possible
configurations are shown in Fig.\ref{fig_three}(a)-(d). First consider the
pure gauge theory, i.e. $m=\infty$. In absence of vertices, there is a
second-order deconfinement phase transition, where the flux tube length
diverges and the quark-antiquark pair loses information about each-other's
position. In presence of vertices, the flux tubes can percolate the space
in a network before the correlation length diverges. That also allows the
quark-antiquark pair (hooked on to the network) to lose information about
each-other's position, but produces a first-order deconfinement phase
transition.

When finite mass quarks are included in the model, they break the flux
tubes by quark-antiquark pair production from the vacuum. With the flux
tube network breaking up, the deconfinement phase transition weakens as
the quark mass is lowered from $m=\infty$
and/or as the chemical potential is increased. Numerical estimates show
that for $N_f=3$ QCD at small chemical potentials, the first-order
deconfinement phase transition ends in a critical line around $m=1.5$GeV,
as depicted in Fig.\ref{fig_one}(a). Although the cross-over region does
not have any sharp signal of the deconfinement phase transition, the
percolating flux tube scenario still suggests a detectable observable there.

It is straightforward to formulate the model on a lattice with spacing $a$.
The flux tubes live on the links, while the quarks and vertices live on
the sites. The variables take values $0,\pm1$, depending on the direction
of the flux, as shown in Fig.\ref{fig_three}(e). The total energy of a flux
tube configuration is
\begin{equation}
E = \sigma a \sum_{i,\mu} |n_{i,\mu}| + m \sum_{i,f} |p_{i,f}|
  + v \sum_i |q_i| ~,
\label{totalenergy}
\end{equation}
where $v$ denotes the energy cost of a vertex, and $f$ sums over the $2N_f$
spin and flavour quark degrees of freedom. The constraint of Gauss's law at
every site is
\begin{equation}
\sum_\mu (n_{i,\mu} - n_{i-\mu,\mu}) - \sum_f p_{i,f} + 3 q_i
\equiv \alpha_i = 0 ~.
\label{gausslaw}
\end{equation}

The grand canonical partition function for the system, with baryon number
$B$, is
\begin{equation}
Z[T,\mu] = \sum_{n_{i,\mu},~p_{i,f},~q_i} \exp\left[
         - {E - 3\mu B \over T} \right] ~ \prod_i \delta_{\alpha_i,0} ~,\quad
B = {1 \over 3} \sum_{i,f} p_{i,f} = \sum_i q_i ~.
\label{partfn1}
\end{equation}
It is fully factorised by expressing the constraint at every site as
$\delta_{\alpha_i,0} = \int_{-\pi}^\pi (d\theta_i/2\pi)
                     e^{i\alpha_i\theta_i}$.
The sum over the variables $n_{i,\mu},~p_{i,f},~q_i$ can then be carried out
explicitly, resulting in
\begin{eqnarray}
Z[T,\mu] &=& \int_{-\pi}^\pi \prod_i {d\theta_i \over 2\pi} ~
             \prod_{i,\mu} (1 + 2e^{-\sigma a/T} \cos(\theta_{i+\mu}-\theta_i))
             \nonumber \\
    &\times& \prod_i \left(1 + 2e^{-m/T}
                     \cos\left(\theta_i + i{\mu\over T}\right)\right)^{2N_f}
    ~\times~ \prod_i (1 + 2e^{-v/T} \cos(3\theta_i)) ~.
\label{partfn2}
\end{eqnarray}
This is in the universality class of the 3-dim XY spin model, in the presence
of an ordinary magnetic field as well as a $Z_3$ symmetric magnetic field.
In this form, $Z[T,\mu]$ possesses a generalised $\cal{PT}$ symmetry,
with $\cal{P}$ corresponding to $\theta_i\rightarrow-\theta_i$ and $\cal{T}$
corresponding to complex conjugation. As a consequence, even though
Eq.(\ref{partfn2}) involves complex weights, the equivalent form in
Eq.(\ref{partfn1}) can be numerically simulated without any fermion sign
problem at finite chemical potential \cite{gattringer,ogilvie}.

Physical interpretation of the site variables $\theta_i$ is uncovered by
insertion of static sources in the system. A static quark at site $j$
modifies the Gauss's law constraint there as
$\delta_{\alpha_j,0} \rightarrow \delta_{\alpha_j,-1}$, and its free
energy is given by $\exp(-F_q/T) = \langle \exp(-i\theta_j) \rangle$.
In the gauge field theory language, this quantity is the expectation
value $\langle P_j \rangle$ of the Polyakov loop at site $j$. So we
infer that $\theta_j$ represents the phase of the Polyakov loop $P_j$,
and the flux tube description of deconfinement is dual to the familiar
Polyakov loop description of deconfinement. The benefit of the flux tube
description is the visual representation it provides of what happens in
position space as the temperature is varied.

\section{Predicted Signal in Heavy Ion Collisions}

Let us consider the flux tube scenario as the fireball of QGP in heavy ion
experiments expands and cools. The cross-over region does not possess a
single percolating flux tube network, but it can still contain many finite
clusters of flux tubes. Since the flux is directed, an obvious feature of
every cluster is that any neighbour of a vertex is an antivertex and vice
versa (see Fig.\ref{fig_three}(d)). This is a topological feature, rather
immune to the details of QCD dynamics. Conservation of baryon number implies
that the vertices can only be locally pair-produced or pair-annihilated.
As the QGP hadronises, the flux tube clusters start breaking up. After the
chemical freeze-out stage, there is no more production or annihilation of
vertices; every vertex ends up in a baryon and every antivertex ends up
in an antibaryon. In the absence of subsequent large-scale diffusion, the
radial propagation of (anti)baryons preserves the geometric pattern of
(anti)vertices present at the chemical freeze-out stage, and the angular
positions of the (anti)baryons seen in the detector can be backtracked to
the angular positions of the (anti)vertices at the chemical freeze-out
stage. This pattern of vertices can then be analysed for correlations
and fluctuations, using techniques similar to those used to analyse the
temperature correlations and fluctuations in the CMBR \cite{weinberg}.

Numerical simulations can calculate the equilibrium correlations between
flux tube vertices in the 3-dim position space, which can then be projected
onto the observed surface of the fireball. Flux tube clusters are bipartite
graphs of degree 3. Their alternating neighbour pattern of vertices and
antivertices can yield two-point baryon number correlations similar
to the oscillatory two-point charge correlations in ionic liquids,
illustrated in Fig.\ref{fig_one}(b). The key common ingredient in the
two cases is the hard-core repulsion between the objects involved.

An oscillatory behaviour of the two-point baryon number correlations
requires the following (in a mimicry of the Sakharov conditions):
(1) A mechanism for local baryon-antibaryon pair production must exist.
This is inherent in the QCD dynamics.
(2) The system must have a non-zero chemical potential (analogous to a
Fermi surface in condensed matter systems). This avoids spectral positivity
constraints, as in case of $\cal{PT}$ symmetric systems \cite{ogilvie},
and is true in experiments.
(3) The dynamical evolution must not be in equilibrium. This favours
fragmentation of flux tube clusters during hadronisation while suppressing
vertex-antivertex annihilation. Production of a sizeable number of
antibaryons in experiments \cite{expt}, from an initial state that has
none, confirms it.

For concreteness, consider a homogeneous and isotropic fluid consisting
of discrete objects with charges $q_i=\pm1$ at positions $r_i$. Then the
average density $\rho_q$ is independent of the position and the average
two-point correlation function $g_q(\vec{r}_i,\vec{r}_j)$ depends only on
$r=|\vec{r}_i-\vec{r}_j|$:
\begin{equation}
\rho_q = \Big\langle \sum_{i} q_i ~ \delta(\vec{r}_i) \Big\rangle ~,\qquad
\rho_q ~ g_q(r) = \Big\langle \sum_{i \ne 0} q_i ~
                  \delta(r-|\vec{r}_i-\vec{r}_0|) \Big\rangle ~.
\label{paircorrel}
\end{equation}
Interactions fade away at long distances, and so $g(r\rightarrow\infty)=1$.
For objects with hard-core repulsion, $g(0)=0$, and beyond the hard core
$g(r)$ tends to its asymptotic value exhibiting damped oscillations
\cite{hardcore}. In particular, the distance scale of the oscillations
is determined by the inter-object separation, and the amplitude of the
oscillations is determined by how tightly the objects are packed together.

To quantify the correlation between vertices and antivertices, it is useful
to compare the correlation functions $g_v$ and $g_{|v|}$ on the same data
sets, respectively obtained with and without the factors of $q_i$ in
Eq.(\ref{paircorrel}). They are sketched in Fig.\ref{fig_one}(b), and their
comparison reduces systematic effects arising from varying fireball sizes.
$g_{|v|}(r)$ is rather insensitive to interactions other than the hard core.
Its first peak is the most informative one regarding the equilibrium fluid
properties---with the location, the height and the area under it respectively
giving estimates of the inter-object spacing, the fluid compressibility
and the number of nearest neighbours. On the contrary, successive correlated
neighbours contribute with opposite signs to $g_v(r)$. For a system with no
correlations between vertices, the probability of occurrence of a vertex or
an antivertex at any location is proportional to its overall density, and
$g_v(r)$ behaves the same way as $g_{|v|}(r)$. Thus the contrast between
$g_v(r)$ and $g_{|v|}(r)$, which is maximum at the first peak, is a measure
of the vertex-antivertex correlations.

Note that a break-up of the flux tube clusters smoothens the oscillations
of the two-point correlation functions, similar to what happens when a
fluid is warmed. Still, if large enough flux tube clusters remain in the
QGP, they would contribute to the contrast between $g_v(r)$ and $g_{|v|}(r)$.
So the strength of the contrast measures the extent to which the nearby
deconfinement phase transition influences QGP properties in the cross-over
region. Specifically, the features to be quantified are the distance scale
and the amplitude of the oscillatory signal. The former is expected to be
the inter-baryon separation ($\sim 2$ fm), while the latter indicates how
tightly or softly the QGP is packed.

Projection of the 3-dim correlation functions onto the observed surface
of the fireball smears their oscillatory structure. That is an easily
calculable effect as long as the inter-vertex separation is smaller than
the radius of the fireball \cite{patel2}. More importantly, the angular
coverage of the experimentally observed hadron distributions is limited.
A partial Fourier expansion, based on the rotation symmetry around the
beam axis and parity, helps in the search for correlations by orthogonal
separation of scales. Labeling the angular distributions by the unit
vector $\hat{n}(\theta,\phi)$, we have
\begin{equation}
b(\hat{n}) = {1\over\sqrt{2\pi}} \sum_{\sigma=\pm}
             \sum_{m=-\infty}^{\infty} b_m^{\sigma}(\theta) ~ e^{im\phi} ~,\quad
\langle b(\hat{n}) b(\hat{n}') \rangle = {1\over2\pi}
             \sum_{\sigma=\pm} \sum_{m=-\infty}^{\infty}
             C_m^{\sigma}(\theta,\theta') ~ e^{im(\phi-\phi')} ~.
\label{baryoncor}
\end{equation}
These expansions of baryon number distributions can be inverted as
\begin{equation}
C_m^{\sigma}(\theta,\theta') = {1\over2\pi}
            \int_0^{2\pi} \!d\phi \int_0^{2\pi} \!d\phi' ~ e^{im(\phi'-\phi)}
            \langle b^{\sigma}(\hat{n}) b^{\sigma}(\hat{n}') \rangle ~,\quad
b^{\pm}(\hat{n}) = {1\over2}
            \left(b(\hat{n}) \pm b(-\hat{n})\right) ~.
\label{twoptcoeff}
\end{equation}
The real and symmetric coefficient functions $C_m^{\pm}(\theta,\theta')$
contain all the information about the two-point correlation functions,
and their experimental values can be tested against model predictions.

The biggest gap between the theoretical formalism and the experimental
data is the fact that the detectors observe protons and anti-protons but
miss neutrons and anti-neutrons. One has to rely on the isospin symmetry
to assume that the observed subset of (anti)protons provides a faithful
characterisation of the total baryon number distribution. Corrections also
need to be estimated for only approximate equilibration of the fireball,
non-uniformity of the QGP caused by the elliptic flow, baryon number
diffusion subsequent to hadronisation and development of the hard baryon
core during hadronisation. Nevertheless, it is imperative to extract the
two-point baryon number correlations from the experimental data and compare
them to theoretical estimates. That would tell us a lot about where exactly
the QGP lies between the extremes of a rigid crystal and a dilute gas.

\end{document}